RESEARCH ARTICLE

OPEN ACCESS

PEER REVIEWED

# Before and after GDPR: tracking in mobile apps

**Konrad Kollnig** *University of Oxford*
**Reuben Binns** *University of Oxford*
**Max Van Kleek** *University of Oxford*
**Ulrik Lyngs** *University of Oxford*
**Jun Zhao** *University of Oxford*
**Claudine Tinsman** *University of Oxford*
**Nigel Shadbolt** *University of Oxford*

**DOI:** https://doi.org/10.14763/2021.4.1611

**Published:** 21 December 2021
**Received:** 19 July 2021 **Accepted:** 8 November 2021

**Funding:** Konrad Kollnig was funded by the UK Engineering and Physical Sciences Research Council (EPSRC) under grant number EP/R513295/1. Max Van Kleek has been supported by the PETRAS National Centre of Excellence for IoT Systems Cybersecurity, which has been funded by the UK EPSRC under grant number EP/S035362/1. The other authors received no specific grant from any funding agency in the public, commercial, or not-for-profit sectors for this research.
**Competing Interests:** The author has declared that no competing interests exist that have influenced the text.

**Abstract:** Third-party tracking, the collection and sharing of behavioural data about individuals, is a significant and ubiquitous privacy threat in mobile apps. The EU General Data Protection Regulation (GDPR) was introduced in 2018 to protect personal data better, but there exists, thus far, limited empirical evidence about its efficacy. This paper studies tracking in nearly two million Android apps from before and after the introduction of the GDPR. Our analysis suggests that there has been limited change in the presence of third-party tracking in apps, and that the concentration of tracking capabilities among a few large gatekeeper companies persists. However, change might be imminent.

# Introduction

The collection of personal data via mobile apps by companies offering analytics, advertising, and other services, has been identified as a significant and ubiquitous threat to data protection and privacy in recent decades (Binns, Lyngs et al., 2018; Wang et al., 2018; Zimmeck et al., 2019). Unlike first-party tracking, where data is collected by the app developers themselves, such companies engage in ‘third party tracking’ where ‘data regarding a particular user’s activity across multiple distinct contexts’ is retained, used or shared between those contexts (Doty et al. (2019)). For instance, a hotel booking app might share a user’s searches with a third-party tracker, who in turn combines that data with the same user’s behaviour obtained from other apps, in order to form a fine-grained profile of the user.

To give citizens ‘better control over how personal data is handled by companies and public administrations’ (European Commission, 2018), the EU updated its data protection regime with the General Data Protection Regulation (GDPR), brought into force in 2018. This law seeks to address, among other aspects, the risks posed by the widespread collection of personal data collection in apps, on the web, and in other digital contexts, by imposing specific requirements in the context of personal data processing. However, limited empirical evidence exists thus far regarding the effect the GDPR has had on the actual act of third-party tracking in smartphone apps.

In this paper, we examine the Android mobile app ecosystem, which remains the largest smartphone app ecosystem. We compare nearly two million Android apps from the UK app store, from before and after the introduction of the GDPR in 2018, to study how the tracking ecosystem has changed. Our data was collected when the UK was still bound to EU law—during the transition period of the EU-UK Withdrawal Agreement. Specifically, we examine the following three research questions:

- RQ1: How has the distribution of third-party trackers across apps on the Google Play Store changed?
- RQ2: How have the organisations doing the tracking themselves changed, in particular in terms of ownership and jurisdiction of operation?
- RQ3: How has the market concentration in third-party tracking changed?

The aim of these questions is to understand, at a macro-scale, whether the GDPR has thus far had a measurable and material impact on the tracking operations of smartphone data aggregators.

Our analysis suggests that there has been limited change in the presence of third-party tracking in apps, limited changes in ownership and jurisdiction of tracking companies, and that the concentration of tracking capabilities among a few large *gatekeeper* companies persists. However, significant change might be imminent, due to recent changes by gatekeeper companies.

We share our code and data from this research at https://osf.io/35xps/.

## Background

In this section, we review background literature and developments that motivate our study. We first introduce the legal background of data protection law, and relate this to third-party tracking. To provide background on our methodological approach, we then present previous literature on analysing data protection and privacy aspects of apps, in particular third-party tracking.

### General Data Protection Regulation (GDPR)

The GDPR (General Data Protection Regulation) came into force in May 2018 to protect data relating to individuals ('personal data'). It replaced the 1995 Data Protection Directive (DPD) aiming to address 'new challenges for the protection of personal data' brought by '[r]apid technological developments and globalisation', as stated in the preamble of the GDPR.. Like the DPD before it, the GDPR places obligations on organisations who process personal data. Those who decide the means and purposes of such processing are 'data controllers', who are required to have a lawful basis for processing (e.g. consent or legitimate interests) (Article 6), and follow principles of fairness, transparency, purpose limitation, data minimisation, accuracy, security and accountability (Article 5). Those who undertake processes on behalf and under the instruction of data controllers are 'data processors' and have a less extensive set of obligations.

In the context of third-party tracking, the first party (e.g. the app developer) is likely a controller; the third parties may be processors (where they only process data on behalf of the first party, e.g. for app analytics), controllers in their own right (where they use the first-party data for their own purposes such as targeted advertising, improving their machine learning models, etc.), or sometimes both at the same time. While some third parties may present themselves as mere processors in order to avoid the obligations of a controller, recent case law of the Court of Justice of the European Union (CJEU) affirms that the bar may indeed be low enough to qualify many third parties as controllers (or joint controllers where they jointly decide on the purposes and means of processing). This has been confirmed, for instance, in *Case C-49/17, Fashion ID, 2019 ECLI:EU:C:2019:629*, finding that when a website embeds a Facebook 'Like' button, which facilitates third-party tracking, it is a joint controller with Facebook; and *Case C-210/16, Unabhängiges Landeszentrum für Datenschutz Schleswig-Holstein v. Wirtschaftsakademie Schleswig-Holstein GmbH*, where the operator of a Facebook fan page operator was deemed a joint controller.

Another important and relevant element of the data protection regime is the 2009 ePrivacy Directive; this covers the privacy of electronic communications and includes rules on the use of cookies and related tracking technologies. Under Article 5(3) of the ePrivacy Directive, third-party tracking typically requires consent as it involves accessing or storing data that is not strictly necessary for delivering the app or service's functionality on a user's device (Kollnig, Binns et al., 2021). The ePrivacy Directive sits alongside and complements data protection law; it constitutes a *lex specialis*, meaning that, when both the ePrivacy Directive and the GDPR apply in a given situation, the rules of the former will override the latter. This means that even if it might otherwise be lawful to process data in third-party tracking under the GDPR without consent (e.g. using an alternative lawful basis like legitimate interests), the ePrivacy Directive would still require consent. Despite the UK leaving the European Union, both the GDPR and the ePrivacy Directive remain unchanged on the domestic UK statute books (at the time of writing), in the form of the UK GDPR and the Privacy and Electronic Communications Regulations (PECR).

### Key changes under the GDPR

Previous data protection law was conceptually and formally very similar to the GDPR, and therefore the legal status and obligations of third-party trackers have not changed substantially (Lynskey, 2015; Voss, 2017; Binns, Lyngs et al., 2018). However, several changes introduced by the GDPR could be expected to make a difference to the compliance efforts of third-party trackers on the ground. In the context of the compliance practices of third-party tracking, three categories of change are particularly pertinent: 1) stricter data protection standards, 2) new governance and accountability obligations, and 3) improved enforcement mechanisms.

**Stricter data protection standards.** The GDPR sets a higher bar for consent to data processing than the DPD (Articles 2 and 7 DPD; Article 7 GDPR). Under the GDPR, consent needs to be freely given, affirmative, specific, unambiguous, and informed. In the context of third-party tracking, these new consent standards have had the effect of enhancing the existing consent requirements under the aforementioned 2009 ePrivacy Directive. Specifically, the ePrivacy Directive requires user consent, according to the improved consent standards of the GDPR, for storing and accessing information on the user device—a prerequisite for most forms of third-party tracking. Moreover, third-party trackers may now struggle to demonstrate their compliance with this consent requirement, as users confronted with first-party consent dialogues may be overwhelmed with information about the tens or hundreds of other third parties involved, and subjected to deceptive design patterns (Nouwens et al., 2020).

**New governance and accountability obligations.** The GDPR introduces new governance and accountability obligations on data controllers. This includes mandatory breach notifications (Articles 33 and 34), record keeping of processing activities (Article 30), data protection officers at larger companies (Articles 37–39), explicit obligations for data processors (Articles 28 and 29), and data protection impact assessments (Article 35). More generally, the GDPR puts forward the principles of data protection by default and design (Article 25), that shall make data protection an integral part of any personal data processing. These new obligations may be much harder for third-party trackers to meet in practice, for example where record keeping of individuals' consent is impossible due to their technical configuration (Commission Nationale de l'Informatique et des Libertés, 2018).

**Improved enforcement mechanisms.** To ensure compliance, the GDPR enables large fines for violations of data protection provisions, of up to €20 million or up to 4% of total global annual turnover (whichever is higher). Further, the GDPR has global reach:

All companies operating in the EU (even those based outside the EU who are processing EU citizens' data) must comply with it (*Lex loci solutionis*). The law also seeks to reduce legal fragmentation amongst EU member states. As a common legal framework for data protection, the GDPR enables the exchange of personal data across the 27 EU member states, thereby allowing businesses to exchange data supposedly seamlessly. Additionally, the GDPR allows for the propagation of personal data beyond member states to countries designated by the EU Commission to have 'adequate' levels of data protection. These countries currently include the UK, Canada, Japan, New Zealand, and Switzerland.

These new enforcement mechanisms are already used in practice, to reduce the tracking of individuals. The French data protection authority CNIL fined Google multiple times over violations of the GDPR (Commission Nationale de l'Informa-tique et des Libertés, 2019, 2020). It also ruled against the practices of the French advertising company Vectaury (Commission Nationale de l'Informatique et des Libertés, 2018). The UK data protection authority ICO investigated the legality of real-time bidding advertising, and stated that the data protection 'issues will [not] be addressed without intervention' (Information Commissioner's Office, 2019). Indeed, the Belgian data protection regulator has recently concluded that real-time bidding, as it is commonly integrated into websites and apps, is in violation of the GDPR (Irish Council for Civil Liberties (2021)). While the regulatory enforcement of data protection law against tech companies was rare under the DPD (Lynskey, 2019; McIntyre, 2020), this seems to have changed since the GDPR, with regulators targeting both smaller (e.g. Vectaury) and larger (e.g. Google) tracking companies. This may reflect not only the changes in provisions of the GDPR compared to the previous data protection regime, but also the increased powers and budgets of regulators since the introduction of the new law (Massé, 2020).

### Heightened privacy expectations of individuals

A further important aspect, that goes beyond the legal text of the GDPR, are the heightened privacy expectations of individuals. Especially since the Edward Snowden leaks in 2013 and the Facebook-Cambridge Analytica revelations in 2018, the public has been interested in how companies and authorities treat their data (Dencik & Cable, 2017; European Commission, 2016). The Snowden leaks were perceived by many individuals, including parliamentarians, as a key moment. As a consequence, the preamble of the GDPR explicitly states that it seeks to tackle the 'widespread public perception that there are significant risks to the protection of natural persons, in particular with regard to online activity'. Tracking is one such risk (Binns, Lyngs et al., 2018; Wang et al., 2018; Zimmeck et al., 2019), especially since US intelligence agencies can access the data troves of Google and other tech companies. These risks posed by US intelligence have recently made the European Court of Justice restrict the sending of personal data to the US, as part of its *Schrems II* judgement.

### Challenges to the effectiveness of GDPR

There are various aspects that challenge the GDPR's effectiveness in practice. The GDPR has led to a proliferation of deceptive and arguably meaningless consent banners online (Matte et al., 2019; Nouwens et al., 2020). Such banners often violate the strict principles for consent under the GDPR and make users' consent process more complicated than intended by the GDPR's transparency principles (Article 5), but the enforcement of the law remains difficult. While growing, many regulators still operate on tight budgets (Lynskey, 2019; Access Now, 2020), and the *one-stop-shop* principle of the GDPR incentivises tech firms to set up their headquarters in member states with relatively lax enforcement. For instance, the Irish Council for Civil Liberties (2021) recently found that Ireland is the 'bottleneck of GDPR enforcement against Big Tech' because of its failure to resolve most major cases against these tech companies. The Age Appropriate Design Code introduced by the UK ICO in September 2021, as a clarification of the GDPR's requirements for children (GDPR-K) in the UK, made explicit requirements for online tracking of children's data against their best interests. However, proving the (non-)existence of tracking activities and their impact on children is expected to be challenging for both technology innovators and law enforcement.

The GDPR is also *technology-neutral*, which can make it difficult for practitioners to translate the GDPR's requirements into software (Bygrave, 2017; Jasmontaite et al., 2018). Smaller companies that lack sufficient legal expertise or compliance budgets (e.g. independent app developers) struggle to implement the GDPR (Ekambaranathan et al., 2021; Sirur et al., 2018). Furthermore, the GDPR does not contain direct obligations for software developers (Bygrave, 2017), and the allocation of responsibility for data processing remains a topic of contentious debate. This is why Giannopoulou (2020) argued that 'more focus should be placed on the level at which privacy design decisions are truly taken and that is at an infrastructural level currently not taken into consideration within the accountability structure of the GDPR'. Especially in the tracking ecosystem, a small number of tracker companies develop the dominant tracking technologies, and ship these to app developers in the form of pre-made tracking libraries. App developers usually neither have access to the corresponding source code nor have a say in how these technologies are developed, and according to whose interests and values (Ekambaranathan et al., 2021).

Paradoxically, the GDPR might actually contribute to the business models of large *ad tech* companies, by putting a market liberal ideology before the protection of personal data (Daly, 2020; Geradin et al., 2020; Gal & Aviv, 2020). At the same time, there is growing evidence that there are indeed widespread infringements of the GDPR and other data protection laws in the app ecosystem (Reyes et al., 2018; Zimmeck et al., 2019; Kollnig, 2019; Okoyomon et al., 2019; Kollnig, Binns et al., 2021; Kollnig et al., 2022).

#### Summary

The changes under the law, particularly the high potential fines, led many to expect that the GDPR would substantially change invasive data collection practices, including third-party tracking. Even though the key principles of the GDPR are similar to those of the DPD, there is reason to believe that the nature and extent of user tracking in mobile apps may have changed since the enforcement of the GDPR in 2018, in light of increased potential fines and regulatory enforcement, a higher bar for consent (which is necessary for most forms of tracking), and heightened expectations of the public. At the same time, the GDPR is not perfect. There remain various challenges to the law's effectiveness, particularly as to how the law integrates into established software development processes.

Our subsequent empirical investigation is not sufficient to establish whether the GDPR is causally responsible for any changes in third-party tracking, and to what extent. However, if the GDPR has indeed, as many had hoped, tackled excesses of personal data processing, we should expect at least some changes in the distribution, ownership, and concentration of third-party tracking in its wake. Further empirical work would be required to establish a causal relationship between the GDPR and such changes.

## App analysis

To observe whether there have been changes in the mobile third-party tracking ecosystem in the wake of the GDPR, we undertake a large-scale analysis of apps in the Google Play Store. To provide methodological background to this analysis, we review the large body of literature that has analysed data protection and privacy properties of mobile apps, and third-party tracking in particular. There are two main methods for doing so: dynamic and static analysis.

### Dynamic analysis

Dynamic analysis investigates the run-time behaviour of apps by executing them on a real smartphone operating system. Most work in this category focuses on analysing apps' network traffic (Han et al., 2019; Le et al., 2015; Razaghpanah et al., 2018; Ren et al., 2016; Shuba et al., 2018; Song & Hengartner, 2015; Van Kleek et al., 2017).

There has been an increasing focus on regulatory issues over recent years. Reyes et al. (2018) analysed the network traffic of children's apps and found a widespread lack of *verifiable parental consent*, which is required under the Children's Online Privacy Protection Act (COPPA) in the US. Okoyomon et al. (2019) found that many apps fail to disclose their data sharing practices fully in their privacy policies. Kollnig, Binns et al. (2021) observed that most apps on the Google Play Store use third-party tracking, but few retrieve the legally required user consent (less than

3.5%).

Dynamic analysis is relatively simple to do and largely device-independent, but it can easily give incomplete results, if not all privacy-relevant aspects of an app are observed during its execution. It also does not scale well across a large number of apps, because every app must be executed individually.

#### Static analysis

Static analysis dissects the behaviour of apps without executing them. With such an approach, Viennot et al. (2014) analysed more than one million apps from the Google Play Store, and found a widespread presence of third-party tracking (including Google Ads in 36% of apps, Facebook in 12%, and Google Analytics in 10%). Similarly, Binns, Zhao et al. (2018) found in analysing nearly one million Google Play apps that about 90% may share data with Google, and 40% with Facebook. This concentration of tracking activities with massive 'gatekeeper' companies can give deep and unexpected insights into our private lives to those companies. It may also raise concerns in competition and antitrust law (Binns & Bietti, 2020; Lynskey, 2019), given the competitive advantage arising from vast data troves (*Bundeskartellamt*, 2019; Competition and Markets Authority, 2020).

Compared to dynamic analysis, static analysis enables the analysis of apps at much larger scale (often millions instead of thousands of apps), but may suffer from both false positives (e.g. if certain parts of the app are not run in practice, but detected as potentially privacy-invasive by the analysis) and false negatives (e.g. if apps load and execute additional privacy-invasive programme code from an external source at runtime).

## Methodology

Our methodology builds on static analysis to analyse tracking in the app ecosystem at scale. We proceed in four steps: app discovery and download, tracking detection, company resolution, and market concentration analysis. The first two steps replicate the work of Binns, Lyngs et al. (2018) on analysing third-party tracking in nearly one million Android apps in 2017, which analysed apps from the UK app store. Since these authors shared their data and analysis tools publicly, this study can be reproduced. The last step replicates a study by Binns, Zhao et al. (2018) that computed the market concentration of tracking companies from 5,000 apps and 5,000 websites, but at a larger scale. In contrast to these previous studies, we focus on the changes of the tracking ecosystem over time and since the introduction of the GDPR. We summarise the limitations of our study in Section 5 ('Discus-sion').

## App discovery and download

To identify and acquire candidate apps for analysis, we used the same approach as Binns, Lyngs et al. (2018) who used an approach derived from Viennot et al. (2014). First, to discover apps, we used an automated process to query the search auto-complete functionality of the Google Play Store with all alphanumeric strings of up to three characters, which returned popular search terms. We then issued these search terms to the Google Play Store search API, and saved all the apps returned as results. We restricted our search to apps available in the UK region of the Google Play Store. We then downloaded all the apps identified in the discovery phase using the open source software gplaycli (Matlink, 2021). We assume here that the data protection practices of apps listed on the UK store do not substantially differ from those listed on the app stores of other countries that have implemented the GDPR.

## Tracking detection

To detect tracking in apps, we performed an automated scan of apps' *.dex files (corresponding to the compiled application code) to identify all URLs (strings starting with http:// or https://). We then manually cross-referenced all URLs corresponding to hosts occurring in at least 0.1% of apps (in 2017 or in 2020), to verify hosts corresponding to trackers. We used the same definition for a tracker as the previous study: *'a third-party tracker [is] an entity that collects data about users from first-party websites and/or apps, to link such data together to build a profile about the user.'* (Binns, Zhao et al., 2018, p. 9)

Overall, we considered more hosts than in the initial study by Binns, Zhao et al. (2018). These authors considered hosts occurring in at least 0.5% of apps in a set of 5,000

Android apps (compared to 0.1% of one million apps from 2017 and 2020 in our study). We additionally verified that our results held when considering the presence of *tracker libraries* in apps, another metric for tracking commonly studied in the literature. The use of tracking libraries is a common way for app developers to integrate tracking capabilities into their apps, because of the ease of integration. However, the detection of tracking libraries might fail if developers use obfuscation techniques to hide their use of tracker libraries (Ma et al. 2016), or use non-standard ways to integrate tracking into their apps (e.g. linking to a Facebook fan page inside an app). This is why we opt for an analysis of tracker hosts in apps, which is more robust towards code obfuscation and the use of non-standard ways of tracking, and also more easily reproducible than past efforts to detect tracking libraries in spite of code obfuscation.

## Company resolution and X-Ray 2020 data set

In order to answer questions about the market structure of the tracking ecosystem, a method for resolving tracking technologies to particular companies and the relationships between them is required. The tracker ecosystem is made up of a large and diverse set of tracker companies, some of which belong to or get acquired by other tracker companies (Binns & Bietti, 2020). For instance, Verizon Communications sold its subsidiaries Flickr and Tumblr, and restructured its online advertising business, see Figure 1. To understand these diffuse company relations, Binns, Zhao et al. (2018) previously created a database of known tracker companies and their company hierarchies in 2017, based on the analysis of 5,000 Android apps.

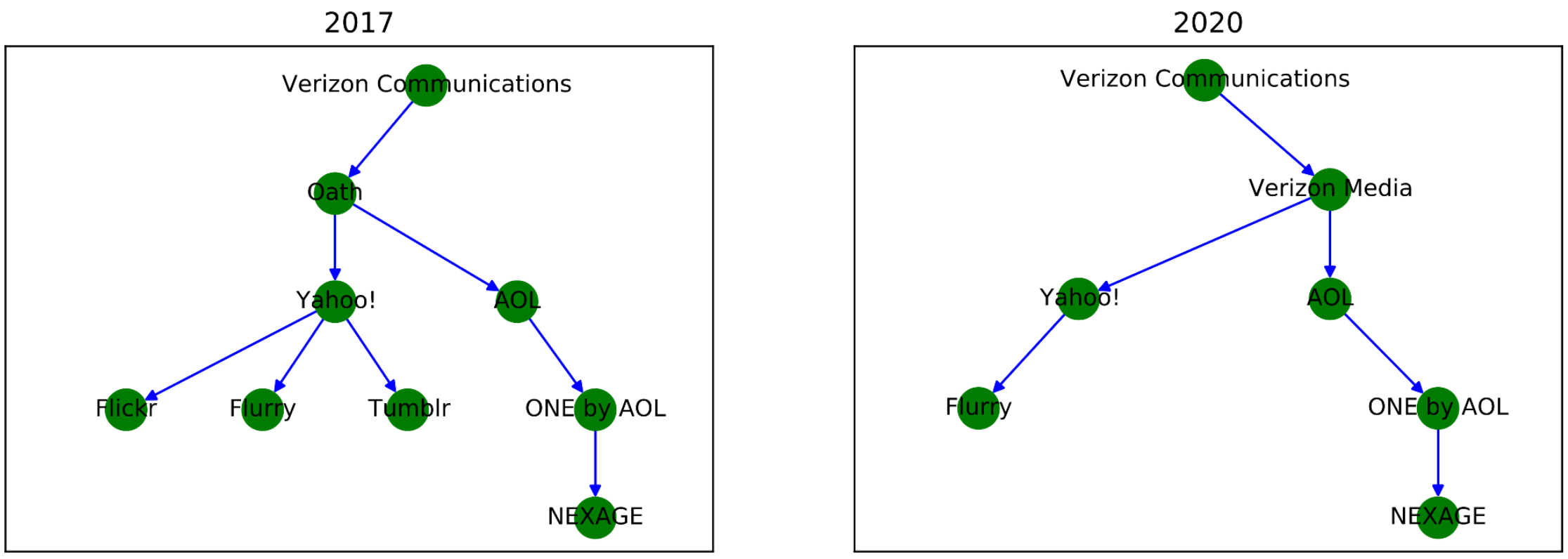

**FIGURE 1:** Company structure of Verizon's tracking business in 2017 and 2020, as an example of the diverse and changing nature of the tracking ecosystem. Only leaf companies present in at least 0.1% of apps are included. 'Verizon Communications' is the *root parent* of the other companies in each graph.

For this study, we created *two* separate tracker databases, one for 2017 and one for 2020 apps. For the 2017 database, we extended the existing database with those tracker hosts additionally found from our analysis, following the same protocols as the previous study. For each new tracker host, we checked to what company it belongs, what parent companies this tracker company has (using WHOIS registration records, Wikipedia, Google, Crunchbase, OpenCorporates, and other public company information), and in what jurisdictions these companies are based. We carefully included only those corporate relations that were already formed by the end of 2017 in the 2017 database.

To create the 2020 database, we revisited every company in our 2017 database, and checked whether its ownership may have changed. We used the same protocols as for the 2017 database to identify what companies are ultimately behind tracking.

Our systematic analysis of tracker libraries and hosts identified 24.4% additional companies (from 578 to 719 companies), comparing our 2017 database to Binns, Zhao et al. (2018)'s original database. Our 2020 database is slightly larger, and contains 754 companies, since it includes additional company transactions that have taken place since 2017. We call the resulting data set X-Ray 2020, and share it with the research community for follow-up studies at https://platformcontrol.org/.

## Market concentration analysis

A common measure for market concentration in economics is the *Herfindahl-Hirschman Index* (HHI). Given market shares $s_1, \ldots, s_N$ for $N$ companies, the HHI is defined as

$$\mathrm{HHI} = \sum_{i=1}^{N} s_i^2. \tag{1}$$

**EQUATION 1:** The *Herfindahl-Hirschman Index* (HHI)

The HHI can attain values between 0 and 1: a higher HHI indicates a more concentrated market. An HHI above 0.1 is considered as potentially concentrated by EU competition regulators (Verouden, 2004), and may motivate a market investigation. US competition regulators use higher thresholds.

The market share of a tracking company is not trivial to investigate. Traditionally, market share is measured in terms of a firm's share of revenue or unit sales of the industry total. However, in the context of free digital services, market share is typically defined in terms of share of users for the service type (e.g. web browsers or search engines which are not revenue-generating or 'sold' to consumers) (Competition and Markets Authority, 2020). Similarly, revenue or unit-sale based measures of market share do not translate into the mobile tracking ecosystem in a straightforward way. Rather, market power in third-party tracking arises from a tracker company's ability to collate personal data across a variety of contexts and generate valuable insights as a result.

To reflect this situation, Binns, Zhao et al. (2018) proposed two measures to measure the market share of a tracker: the integration share (ISH) and the prominence-weighted integration share (PROWISH). The ISH measures the popularity of a tracker with *app developers*, and expresses this popularity relative to other trackers. The PROWISH measures the presence of a tracker in apps most popular with *app users*, also relative to other trackers. Using the ISH (PROWISH) in Equation (1) then gives the ISH-HHI (PROWISH-HHI). For details on the computation, see the Appendix.

The assessment of market shares remains subject of ongoing debate; so far, there has been limited intervention by competition authorities against excessive and increasing access to personal data by a single company (Binns & Bietti, 2020).

# Results

## Downloaded apps, installs, and app death

We downloaded a total of 1,000,750 apps between January and March 2020. This is about 2.5 years after the original study, which collected 958,270 apps between August and September 2017. Only 33.9% of the previous apps were still available on the Google Play Store in 2020; the remaining Play Store entries did not exist anymore (though they might still exist elsewhere, e.g. outside the Play Store). The median app was last updated on the Play Store in January 2017 for the 2017 data set and in June 2019 for the 2020 data set. 75.8% of 2020 apps were last updated since 25 May 2018, when the GDPR came into force.

## Numbers of distinct tracker hosts in apps

Apps from both years contained a high number of distinct hosts in their source code that belong to tracker companies (‘tracker hosts’). Their number was highly right-skewed, see Figure 2 (left). The median number of tracker hosts included in an app was 9 in 2017, and 11 in 2020. 14.30% of 2017 apps contained more than 20 tracker hosts, compared to 15.72% in 2020. 88.44% contained at least one in 2017, and 91.37% in 2020, a slight increase.

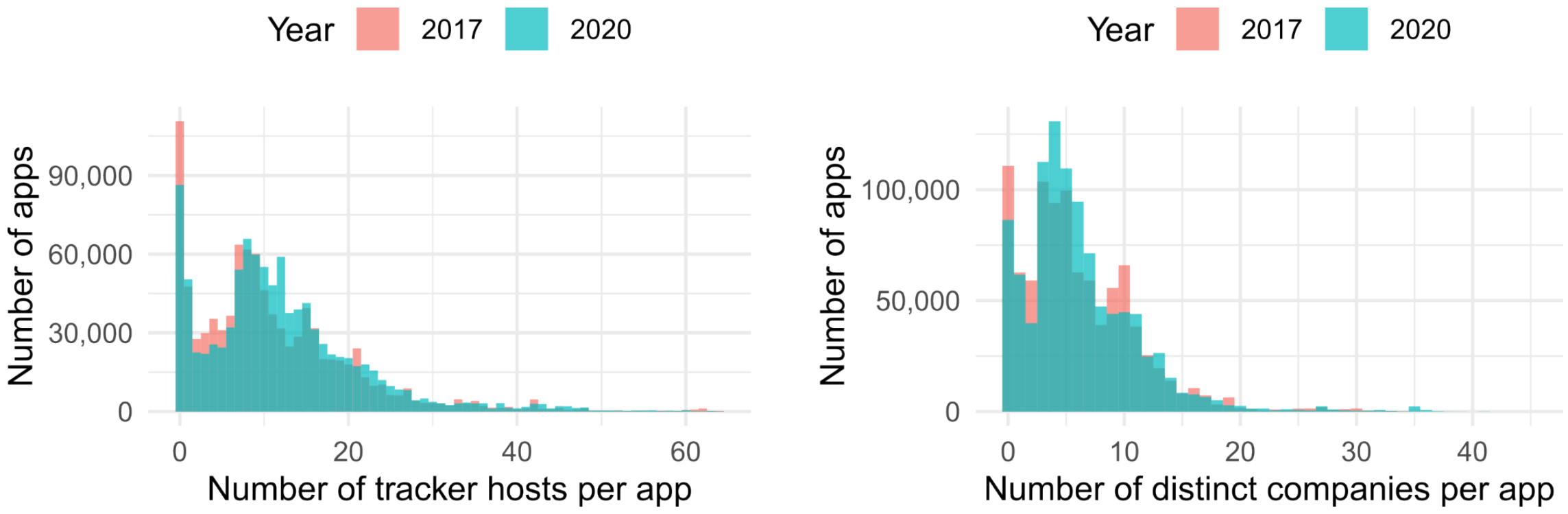

| Year | Median | Count>20 | None |
|---|---|---|---|
| 2017 | 9 | 14.30% | 11.56% |
| 2020 | 11 | 15.72% | 8.63% |

| Year | Median | Count>10 | None |
|---|---|---|---|
| 2017 | 5 | 15.27% | 11.56% |
| 2020 | 5 | 15.76% | 8.63% |

**FIGURE 2**: Number of tracker hosts per app (left) and companies behind hosts (right) in free apps on the Google Play Store. We exclude extreme outliers having more than 65 tracker hosts (left).

### Numbers of distinct companies behind hosts

The prevalence of 'leaf' tracker companies (i.e. companies at the lowest subsidiary level, such as 'Flurry' as a subsidiary of 'Yahoo!' in Figure 1) in apps was highly right-skewed, see Figure 3 (right). The median number of companies was 5 in both years. 15.27% contained more than 10 companies in 2017, 15.76% in 2020.

The maximum number of companies referenced in a single app was 45 in 2017, and 43 in 2020. Amongst the 68 apps from both years that referenced more than 40 companies, 34 were related to photo editing, 21 to dating, 7 to sports news, 2 to games, and 1 to time tracking. This underlines how seemingly innocent apps (e.g. photo editing, time tracking) but also highly sensitive apps (e.g. dating) can expose personal data to an unexpected number of companies.

Since many tracker companies belong to a larger consortium of companies, we can also consider tracking by 'root parent' (e.g. 'Flurry' is ultimately owned by its root parent 'Verizon Communications', see Figure 1). Figure 3 shows both the 'prevalence' of root parents (i.e. the percentage of apps that contain this tracker) and their 'prominence' (i.e. the percentage of total app installs that ship this tracker).

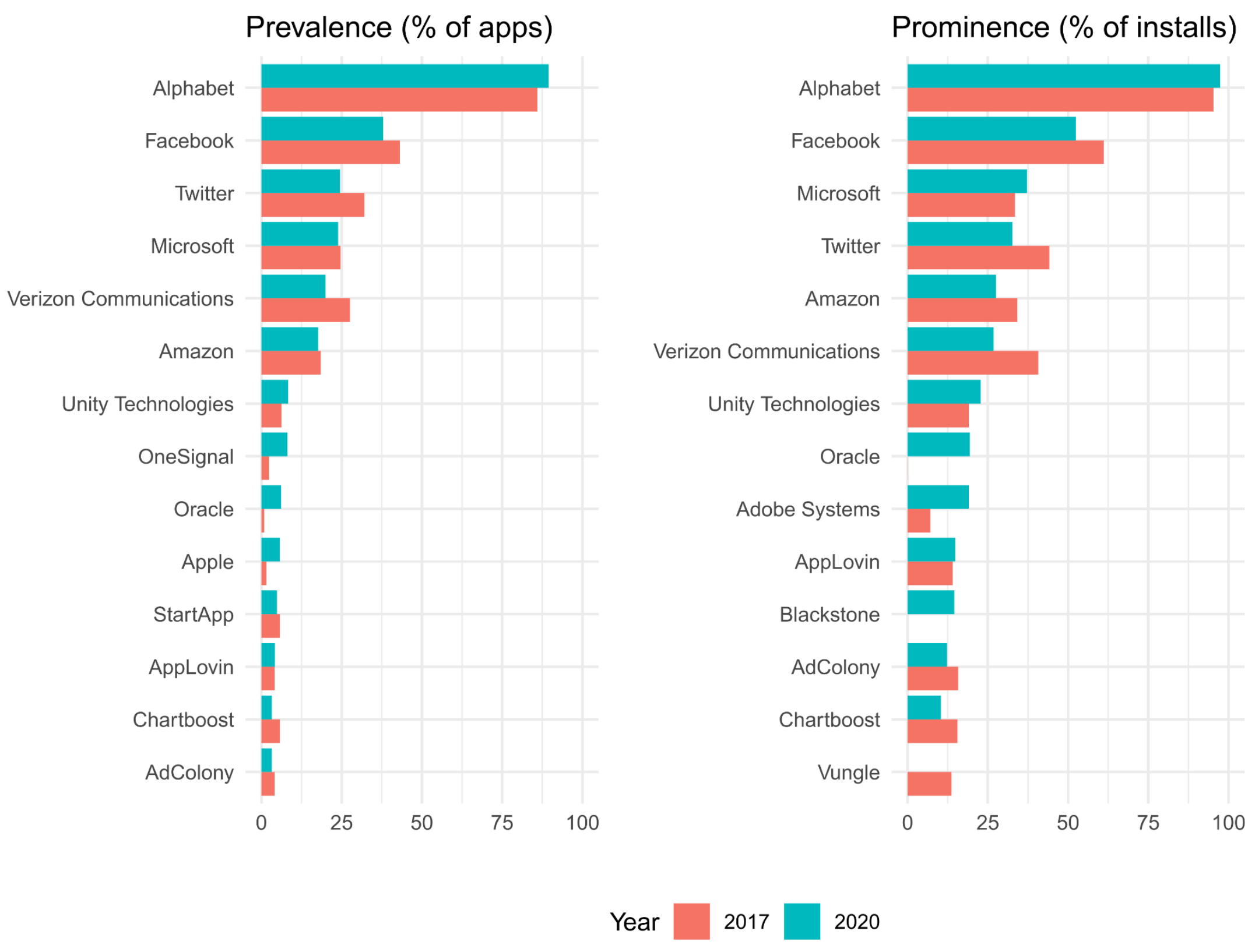

**FIGURE 3**: Prevalence and prominence of hosts relating to certain root tracking companies in apps in 2017 and 2020. We consider the top 11 companies from both years and for both prevalence and prominence. The companies are ranked by the values in 2020.

The overwhelming share of apps included hosts belonging to Alphabet/Google and Meta/Facebook. Alphabet/Google has even increased its presence in apps slightly, while Meta/Facebook has lost some market share. Twitter has also lost some market share, and has been overtaken by Microsoft. Oracle has greatly increased its market share (especially in prominence), since its acquisition of Moat in 2017. Beyond these digital behemoths, many specialised tracking companies (including AppLovin, AdColony, Chartboost) are among the market leaders when considering their 'prominence' (i.e. share of app installs). The prominence plot also reflects the acquisition of Vungle by Blackstone in 2019, resulting in a change of root company. The median tracker company has increased its market share (prevalence and prominence both up from 3.1 in 2017 to 3.4 in 2020). Overall, the tracking market has seen no new entrant into the top 7 companies, both when ranking by 'prevalence' and 'prominence'.

## Company prevalence by genre

There exists a wide range of genres on the Google Play Store to help users explore apps better. The overall number has remained at 49 since 2017. Since the genres have stayed the same, we group these genres into the same 8 'super genres' as the previous paper to provide high-level statistics about the apps (for example, the genres 'Comics', 'Sports', 'Video Players', and all games are all grouped into 'Games & Entertainment'). Children' apps are singled out (which are assigned to 'Family' categories on the Play Store), given the concern around data collection from this group of app users. We re-ran the company analysis for each super genre, see Figure 4.

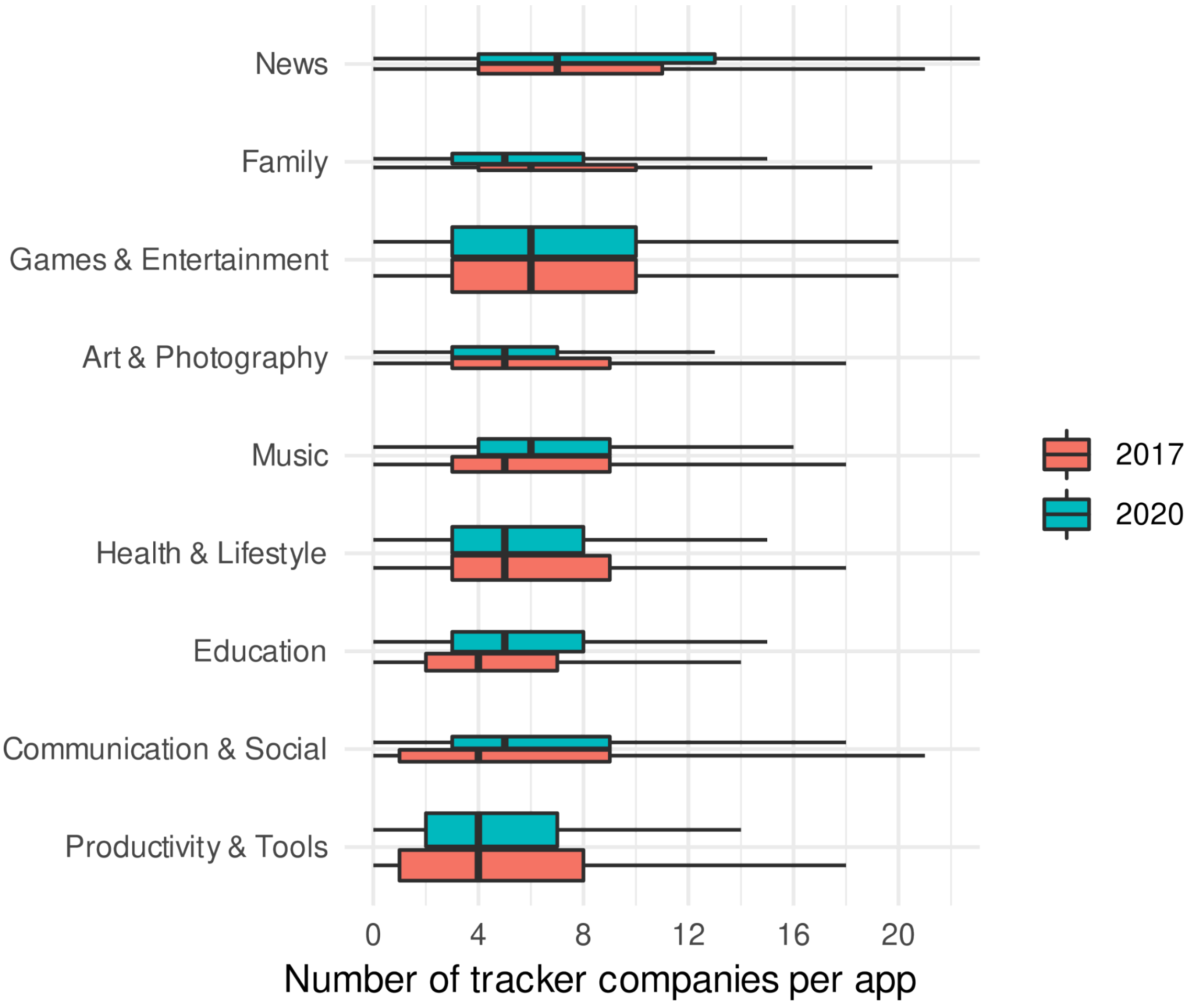

**FIGURE 4:** Boxplot of number of distinct tracker companies behind hosts referenced in apps, grouped by super genre. Black bars indicate medians. Height of bars indicates number of apps in a given super genre.

The genre with most tracker companies was 'News' (seven companies), both in 2017 and 2020. Family (Children apps), having the second most tracker companies in 2017 (6 companies), was down by one company in the median. 'Music', 'Education', 'Communication & Social' were up by one company on average. Overall, the presence of tracking in apps was similar between the years across super genres.

### Country differences

We also analysed in which countries the tracker companies in apps are based (including the subsidiary and all parent companies, see Table 1). About 90% of apps contained a tracker that is owned by a US-based company. The next most common countries were China and Russia in both years. The top 8 countries were also the same, but with different rankings. South Korea has seen an increase by 22%, while Germany and Israel have both seen a decline by about 40%. However, overall, the fluctuation between the years is small across the top 8 countries. Overall, the share of tracker companies based in UK and EU member states has both somewhat decreased between 2017 and 2020.

| Country | % Apps (2017) | % Apps (2020) |
|---|---|---|
| US | 88.38 | 91.31 |
| China | 6.35 | 6.11 |
| Russia | 4.12 | 4.19 |
| Germany | 4.04 | 2.42 |
| South Korea | 3.18 | 3.88 |
| UK | 2.92 | 2.68 |
| India | 2.11 | 1.83 |
| Netherlands | 1.87 | 1.57 |

**TABLE 1:** Apps including at least one tracker associated with a company within a given country.

We also computed the country prevalence for each super genre. The US stayed the most prevalent country (between 83% and 95% for 2017, and 88% and 97% for 2020). China was present in about 9.2% of apps from the 'Health & Lifestyle' genre in both years.

## Changes in company structure

We now analyse the network of companies involved in tracking, and how this has changed from 2017 to 2020. We included those tracker hosts occurring in more than 0.1% of apps, the tracking companies owning these hosts, and all their parent companies. We refer to all included companies as having a *significant market share* in app tracking. 0.1% might seem small, but can still amount to millions of individuals, since there are billions of Android users.

In total, there were 164 companies (including all parent companies) with a significant market share in 2017, compared to 162 in 2020. There were a total of 102 root companies with a significant market share in 2017, compared to 89 in 2020. On average, a company consortium consisted of 1.48 companies in 2017, compared to 1.65 in 2020. 62 companies were owned by another in 2017, compared to 73 in 2017. All these figures point to a subtle consolidation of tracking companies since 2017.

One straightforward explanation for the consolidation of the tracking ecosystem would be a substantial number of companies *losing a previously significant market share* (i.e. losing access to at least 0.1% of apps). Companies that have lost their significant market share include Myspace, Loggly, and BugSense. However, no larger tracking companies have been affected by this.

Another important reason for consolidation in the tracking market were *mergers and acquisitions (M&A)*. We found a total of 53 M&A transactions between the beginning of 2018 and June 2020 among tracker companies. For instance, Blackstone, one of the largest investment firms, entered the tracking market with the purchase of the advertising firm Vungle in July 2019. Media Games Invest, another investment firm, purchased the tracking companies PubNative and Verve, as part of over 30 strategic acquisitions over the past six years (Gardt, 2020). Verve, in turn, had purchased the advertising company Receptiv in May 2018. Overall, there were 7 investment firms in our company data set with a significant market share.

Three of the 53 observed M&A transactions were filed with EU or UK competition authorities: Bain Capital Investors / Kantar, Silver Lake / ZPG, and Taboola / Outbrain. The first two were filed with the European Commission, which did not pursue in-depth investigations and approved the M&A transactions within a few weeks. Taboola filed the planned acquisition of its rival Outbrain with the UK Competition and Markets Authority (CMA) in April 2020. The CMA opened a phase 1 investigation, and found potential competition concerns leading to a phase 2 investigation from June 2020. Taboola eventually abandoned its acquisition plans in September 2020, which made the CMA cancel its investigations.

We have also observed 11 rebrandings among prevalent tracking companies. For example, Verizon Communications has restructured its media operations internally, inside its subsidiary Verizon Media (previously known as Oath). Amazon renamed its 'Amazon Marketing Services' to 'Amazon Advertising', thereby seemingly trying to advance its mobile advertising business. Amazon also purchased the advertising firm Sizmek in 2019, after a three-year ownership by the private equity firm Vector Capital led to bankruptcy. Microsoft has integrated BitStadium (purchased in 2014) into its other cloud services, and rebranded it as 'Microsoft App Center'. Notably, after our data collection, Facebook rebranded itself as 'Meta'.

### Market concentration

We now consider how the market concentration of tracking companies has changed between 2017 and 2020. As discussed in our methodology in Section 3, we use two metrics: the ISH-HHI and PROWISH-HHI. The results can be seen in Table 2.

| Year | ISH-HHI | PROWISH-HHI | Gini |
|---|---|---|---|
| 2017 | 0.112 | 0.071 | 0.491 |
| 2020 | 0.115 | 0.067 | 0.493 |

**TABLE 2:** Market concentration and equality measures. All three metrics take values between 0 and 1. A higher HHI value indicates more concentration in the tracking market. A lower Gini coefficient indicates higher equality between the market participants.

The ISH-HHI has seen a subtle increase, the PROWISH-HHI a subtle decrease. Since the ISH-HHI is in the range of 0.1, this shows some signs of concentration in the integration of tracking into all apps in both years. However, when weighting apps by their prominence (i.e. by number of app installs), the concentration decreases to about 0.07. The Gini coefficient, an inequality metric herein computed amongst root tracking companies, has increased subtly (see Table 2). This suggests a slightly decreased equality in terms of market access of tracker companies in 2020.

Overall, there has been very limited change across all studied market concentration measures.

## Discussion

In this section, we discuss the above findings in the context of our original re-search questions, and their implications for the ongoing development of data rights regulation and the future of the third-party tracking sector.

**The distribution of third-party trackers has not changed much.** Our results suggest that the GDPR has not had a large effect on the distribution of third-party tracking across apps on the UK Google Play Store. The same handful of third-party tracking companies have similar prevalence and prominence; the average app contains a similar number of third-party trackers (measured at the level of companies rather than hosts); and a consistent percentage of apps (15%) contain more than ten trackers. If the GDPR has led to changes in tracking practices, they are not showing in the distribution of trackers. This might seem surprising, given that the GDPR and ePrivacy present challenges for compliance in the context of multiple third parties. Rather than reduce the number of third parties they share data with, to enable compliance with the requirements of consent, record-keeping, data protection by design, transparency and accountability, first-party app developers continue to share data with multiple third parties.

Some small changes in the distribution of third-party trackers have been observed. Alphabet-owned trackers have slightly increased in both prevalence and prominence, while others such as Meta/Facebook and Twitter have decreased on both measures. The number of apps with no trackers at all has decreased from 11.6% to 8.6%. While these might in some way be indirect effects of the GDPR, we find no clear explanation connecting them.

**Cross-jurisdictional data flows.** As explained above, the EU data protection regime enables the free-flow of personal data across EU member states, the UK and other countries that are deemed to provide 'adequate' data protection standards, as determined by the European Commission. This does not mean that data being sent to a third-party tracker based outside the EU / UK's list of adequate countries is necessarily unlawful; some tracker companies may designate local subsidiaries as the data controller for personal data of citizens in the EU / UK, and transfers to non-adequate countries may still be lawful with the use of alternative measures including 'standard contractual clauses' (Article 46 GDPR) and 'binding corporate rules' (Article 47 GDPR).

Given that these alternative options come at substantial costs, it would be reasonable to expect at least some decrease in the number of third-party trackers based in non-adequate countries, as first parties seek to minimise the compliance risk of unlawfully transferring data across borders. However, despite their jurisdiction not being deemed as 'adequate', companies based in the US, India, China, Russia were still behind a large portion of the tracking observed in our analysis in 2020. In particular, organisations based in the US (about 90%) and China (about 9%) led the pack for third-party tracking in the 'Health & Lifestyle' super genre. These findings are potentially concerning: Absent specific justifications, the GDPR prohibits processing data concerning health (see Article 9(1) GDPR). While our study did not determine which, if any, third-party trackers were collecting data that could be treated as health-related data under the GDPR, there is often a risk of accidental disclosure of sensitive information (e.g. the information that an individual uses certain sobriety or mental health apps) (Norwegian Consumer Council, 2020). Overall, there has been limited change in sending data to trackers in non-adequate countries (which includes the US). Indeed, there is a slight reduction in the prevalence of third-party trackers based in significant countries *inside* the EU (Germany and the Netherlands), supporting the claim that GDPR may actually be helping global tech firms *outside* the EU (Geradin, 2020).

**Market concentration and competition.** Our analysis hints at a high level of concentration in the tracking market. Alphabet/Google and Meta/Facebook continue to dominate app tracking. Their dominance is particularly present in the number of apps they cover. If these companies can show ads on devices that other competitor advertising companies hardly have access to (e.g. due to the *default bias* of app developers to use the software solutions of established brands, see Ekambaranathan et al. (2021), Mhaidli et al. (2019)), they can extract sizeable revenues from their dominance of the tracking market, and might even be able to exert meaningful control over advertising prices (Competition and Markets Authority, 2020). From our data, this seems to be particularly the case for those apps that have few installs, but represent the vast majority of apps on the Play Store due to the long-tailed distribution (Binns, Lyngs et al., 2018; Viennot et al., 2014).

At the same time, many relatively smaller companies are involved in app tracking. Some of these manage to reach fairly high market shares in terms of app installs (including AppLovin, AdColony, and Chartboost). These smaller companies usually focus exclusively on mobile advertising, instead of having a broad portfolio of digital services like Alphabet/Google, Meta/Facebook, or Verizon. The specialisation and small size of these tracking companies seems to allow them to gain a certain competitive advantage, and potentially offer better deals to app publishers (who might otherwise choose the market leaders). An important competitive advantage of these companies might be reduced public awareness and regulatory scrutiny, allowing them to compete with the market leaders in certain segments, at the expense of data protection and user privacy.

Smaller companies may have access to fewer apps, but they might still be able to gain deep insights into the lives of individuals, especially at the aggregate level. Even if a tracker company gets access to a small subset of users only, the use of *permanent user identifiers* can enable these companies to exchange data with other tracking companies, such as data brokers, and gain insights into larger numbers of users. The average user has about 30 apps installed at any given time (Audience-Project, 2020; Google, 2016), but for a third-party aiming to obtain a profile of the user, it might be sufficient to be integrated into only one of those apps. As such, there may be diminishing returns for third parties aiming to increase their prevalence or prominence in the app marketplace.

While a concentration of data with only a few companies can help transparency of tracking and compliance with data protection and privacy legislation, it also puts more power into the hands of a few companies. By contrast, a tracking ecosystem with dozens of market participants—as we continue to have—is difficult to oversee by regulators and the interested public.

**Limitations.** Our work has certain limitations. The analysis of hosts in apps only gives a partial picture of app tracking, as explained in Section 3.2. We do not analyse the handling of personal data on the servers of tracking companies or how these companies might share data with other companies, but only tracking that happens directly on users' devices. We only focus on tracker hosts that are present in at least 0.1% of apps. Some of these hosts may never be contacted, while other hosts may not be present in the app code at install time. Further, the definition of 'tracking' (see Section 3.2) is, while based on the protocols of previous research by Binns, Zhao et al. (2018), open to debate. Lastly, we treat all tracker hosts equally, and do not account for different purposes (e.g. advertising and analytics) or for different levels of intrusiveness. While this paper focuses on Android apps and the Google Play Store, tracking is also widespread on iOS and the Apple App Store (J. Han et al., 2013; Kollnig et al., 2022).

# Conclusions

In this work, we analysed the presence of third-party tracking in apps, before and after the introduction of the GDPR. Most instances of third-party tracking without user consent were already against the law before the GDPR, under the 2009 ePrivacy Directive; the GDPR has tightened the previous rules around consent even more. Our analysis shows that the number of third-party tracking services integrated into mobile apps has not changed massively, and that most apps still integrate third-party tracking when privacy-preserving alternatives (e.g. Matomo, ACRA, con-textual ads) exist. Previous research, analysing a representative subset of the same 2020 app data set as in this paper, found that about 70% of apps sent data to tracking companies immediately at the first app start; less than 10% asked for the legally required consent (Kollnig, Binns et al., 2021). The Schrems II Ruling from July 2020 (i.e. after our data collection) has also outlawed most transmissions of personal data to the US; we do not expect that this has changed the integration of the dominant tracking services by Alphabet/Google and Meta/Facebook into apps, and the subsequent sending of personal data to the US. In conclusion, while the GDPR itself is strict around third-party tracking, compliance with basic provisions of the law in apps on the Google Play Store is limited. This suggests widespread infringements of the GDPR, as well as an ongoing lack of enforcement of data protection rules on the Google Play Store.

Specifically, we found that tracking has remained prevalent across a wide range of mobile apps and prominent in its reach of app user data. The number of tracking companies has stayed about the same between 2017 and 2020 in the average app on Google Play. The top destination countries have likewise stayed the same, as have the most prominent tracking companies—namely Alphabet/Google and Meta/Facebook—and the sending of personal data to trackers based in a third-party country without an 'adequate' level of data protection. Our observations are consistent across *super genres*. Apps continue to rely on tracking technologies, e.g. to retrieve analytics and show advertising, even after the introduction of the GDPR. The law does not appear to have changed these incentive structures fundamentally.

We also found that the market concentration in the tracking ecosystem has seen limited change over time. Competition between tracking companies seems to revolve at least partly around data protection and user privacy due to the relevance of little-known tracking companies that evade public and regulatory scrutiny but collect data about sizeable numbers of individuals. As such, our study provides empirical evidence of fears expressed in previous academic work that the GDPR might entrench the existing power imbalances in the digital ecosystem (Daly, 2020; Geradin et al., 2020; Gal & Aviv, 2020).

While our current analysis points to limited change in the tracking ecosystem so far, change might be imminent. Apple and Google have been introducing various privacy measures that could, despite increasing the concentration of data collection with these companies, improve data protection and user privacy. The most notable recent example is Apple's introduction of mandatory user opt-ins to tracking in iOS apps in April 2021. First reports suggest high refusal rates of tracking (Rosenfelder, 2021; Flurry, 2021), with the direct result of tripling the iOS market share of Apple's own advertising business (Financial Times, 2021), which itself sidesteps the new rules against tracking (Seufert, 2021). However, the effects of this new policy are still subject to ongoing debate and analysis. Meanwhile, Google is considering removing third-party cookies from its Google Chrome browser and replacing them with Federated Learning of Cohorts (FLOC), thereby shifting away from identifying individuals to targeting cohorts of users with similar interests.

An important driver of these new privacy measures has been the emergence and overhaul of data protection and privacy laws around the globe, more extensive regulatory action, and ultimately the increased privacy expectations of citizens. In this sense, the GDPR has already contributed to changing the mobile tracking ecosystem by shaping people's expectations around privacy and increasing data protection enforcement. Beyond the EU, the GDPR has also encouraged the emergence of new and revised data protection laws, notably in Brazil, Japan, China and California. In the UK, the government is discussing a reform of its domestic implementation of the GDPR. Meanwhile, the EU is planning to introduce a new ePrivacy Regulation, which would overhaul and supersede the existing ePrivacy Directive in the EU, but not in the UK, leading to further regulatory divergence. According to our analysis, the lack of enforcement of the existing rules is one of the key issues that needs to be addressed.

Transparency is essential in keeping power to account, but the analysis of privacy practices remains difficult in the mobile tracking ecosystem. This conflicts with the strict transparency requirements for the processing of personal data laid out in the GDPR (Article 5). As a result, we only analyse the presence of tracking in apps, but not the handling of data behind the scenes by those third parties beyond their initial data collection. More research as well as changes to the current data protection and privacy practices of the gatekeepers will be needed to afford regulators and independent researchers more transparent access and to build more sustainable business models that can live without the continuous surveillance of those individuals that these technologies are meant to serve.

# Appendix

## Computation of market concentration measures

The integration share (ISH) $s_i$ of a tracker company $t_i$ is computed from its prevalence, i.e. the percentage of apps that this company is present in:

$$s_i = \frac{\mathrm{prevalence}(t_i)}{\sum_{j=1}^{N} \mathrm{prevalence}(t_j)}. \tag{2}$$

**EQUATION 2:** The integration share (ISH) $s_i$ of a tracker company $t_i$

The prevalence has been used widely across in the app analysis literature, as a means to assess tracker adoption in apps. Computing the ISH $s_i$ for every tracker company $t_i$, and using the computed $s_i$ in Equation (1) yields the ISH-HHI.

We additionally study the prominence of a tracker company $t_i$ as the share of overall app installs that this company is present in:

$$\mathrm{prominence}(t_i) = \frac{\text{Sum of installs of apps with tracker } t_i}{\text{Sum of all app installs}} \tag{3}$$

**EQUATION 3:** The prominence of a tracker company $t_i$ as the share of overall app installs

Using the prominence instead of the prevalence in Equation (2) gives a prominence-weighted integration share (PROWISH) $s_i$. Using the $s_i$ computed from the prominence in Equation (1) gives the PROWISH-HHI. For a more in-depth discussion, see the original publication by Binns, Zhao et al. (2018).

It is important to note that we compute the PROWISH differently than in previous work (Binns, Zhao et al., 2018; Englehardt & Narayanan, 2016), which focused on app ranks instead of app installs. The aim of both approaches is the same: approximate the number of users that a tracker company has access to.

# Detailed comparison of companies behind tracker hosts in apps

(a) 2017

| Root | % apps | Subsidiary | % apps | Country |
|---|---|---|---|---|
| Alphabet | 86.04 | Google | 84.75 | US |
| | | Google APIs | 67.77 | US |
| | | Analytics | 40.6 | US |
| | | Tag Manager | 36.98 | US |
| | | AdSense | 30.48 | US |
| | | Firebase | 21.62 | US |
| | | AdMob | 14.33 | US |
| | | YouTube | 10.06 | US |
| | | Crashlytics | 7.33 | US |
| | | Fabric | 5.87 | US |
| | | Blogger | 0.35 | US |
| | | Plus Codes | 0.08 | US |
| | | DoubleClick | 0.07 | US |
| | | Dialogflow | 0.01 | US |
| Facebook | 43.14 | Facebook | 43.01 | US |
| | | Instagram | 2.22 | US |
| | | LiveRail | 1.2 | US |
| | | Account Kit | 0.07 | US |
| Twitter | 32.05 | Twitter | 31.87 | US |
| | | MoPub | 2.79 | US |
| Verizon | 27.61 | Yahoo! | 22.07 | US |
| | | Flurry | 6.57 | US |
| | | Tumblr | 2.08 | US |
| | | Flickr | 1.95 | US |
| | | ONE by AOL | 0.73 | US |
| | | NEXAGE | 0.37 | US |
| | | AOL | 0.04 | US |
| | | Jumptap | 0.02 | US |
| | | BrightRoll | <0.01 | US |
| | | Gravity | <0.01 | US |
| Microsoft | 24.64 | Microsoft | 23.33 | US |
| | | LinkedIn | 21.79 | US |
| | | Bit Stadium | 0.78 | DE |
| | | Yammer | 0.51 | US |
| | | Virtual Earth | 0.15 | US |
| | | Bing | 0.13 | US |
| | | App Center | 0.07 | US |
| Amazon | 18.49 | Web Services | 11.93 | US |
| | | Amazon | 8.22 | US |
| | | Marketing Services | 1.78 | US |
| | | Alexa | <0.01 | US |

(b) 2020

| Root | % apps | Subsidiary | % apps | Country |
|---|---|---|---|---|
| Alphabet | 89.49 | Google | 88.83 | US |
| | | Google APIs | 77.68 | US |
| | | AdSense | 66.1 | US |
| | | Firebase | 39.88 | US |
| | | Analytics | 20.67 | US |
| | | Crashlytics | 16.16 | US |
| | | Tag Manager | 15.82 | US |
| | | AdMob | 12.05 | US |
| | | YouTube | 10.18 | US |
| | | Fabric | 9.53 | US |
| | | Blogger | 1.82 | US |
| | | Dialogflow | 1.36 | US |
| | | Plus Codes | 0.2 | US |
| | | DoubleClick | 0.02 | US |
| Facebook | 37.91 | Facebook | 37.33 | US |
| | | Instagram | 3.57 | US |
| | | Account Kit | 0.25 | US |
| | | LiveRail | 0.2 | US |
| | | WhatsApp | 0.15 | US |
| Twitter | 24.5 | Twitter | 24.44 | US |
| | | MoPub | 2.05 | US |
| Microsoft | 23.9 | Microsoft | 22.13 | US |
| | | LinkedIn | 15.2 | US |
| | | App Center | 1.47 | US |
| | | Bit Stadium | 0.72 | DE |
| | | Yammer | 0.26 | US |
| | | Bing | 0.19 | US |
| | | Virtual Earth | 0.16 | US |
| | | Drawbridge | <0.01 | US |
| Verizon | 19.93 | Yahoo! | 15.83 | US |
| | | Flurry | 5.85 | US |
| | | Verizon Media | 0.59 | US |
| | | NEXAGE | 0.18 | US |
| | | ONE by AOL | 0.12 | US |
| | | AOL | 0.03 | US |
| | | Jumptap | <0.01 | US |
| | | BrightRoll | <0.01 | US |
| Amazon | 17.68 | Web Services | 10.07 | US |
| | | Amazon | 9.14 | US |
| | | Advertising | 1.03 | US |
| | | Alexa | <0.01 | US |

**TABLE 3:** Top tracker companies, and their subsidiaries, derived from analysing tracker hosts in apps